\documentclass[%
 preprint,
nofootinbib,
showkeys,
 amsmath,amssymb,
 aps,
 superscriptaddress,
]{revtex4-2}

\usepackage{dcolumn}
\usepackage{bm}

\usepackage{amsmath}
\usepackage{amssymb}
\usepackage{ragged2e}
\usepackage{xcolor}
\usepackage{graphicx}
\usepackage[ruled,lined]{algorithm2e}

\SetCommentSty{mycommfont}

\usepackage{appendix}
\usepackage{orcidlink}

\newcommand{\abs}[1]{\left\vert #1 \right\vert}
\newcommand{\qbar}{\overline{q}}
\newcommand{\ombar}{\overline{\omega}}
\newcommand{\betabar}{\overline{\beta}}
\newcommand{\mubar}{\overline{\mu}}

\begin{document}

\preprint{APS/123-QED}

\title{Monte-Carlo Event Generation for X-Ray Thomson Scattering Analysis}

\author{Uwe Hernandez Acosta \orcidlink{0000-0002-6182-1481}}
\email[Corresponding author: ]{u.hernandez@hzdr.de}
\affiliation{Center for Advanced Systems Understanding (CASUS), Untermarkt 20, 02826 Görlitz, Germany}
\affiliation{Helmholtz-Zentrum Dresden-Rossendorf (HZDR), Bautzner Landstraße 400, 01328 Dresden, Germany}
\author{Tomas Gawne}
\affiliation{Center for Advanced Systems Understanding (CASUS), Untermarkt 20, 02826 Görlitz, Germany}
\affiliation{Helmholtz-Zentrum Dresden-Rossendorf (HZDR), Bautzner Landstraße 400, 01328 Dresden, Germany}
\author{Jan Vorberger}
\affiliation{Helmholtz-Zentrum Dresden-Rossendorf (HZDR), Bautzner Landstraße 400, 01328 Dresden, Germany}
\author{Hannah Bellenbaum}
\affiliation{Center for Advanced Systems Understanding (CASUS), Untermarkt 20, 02826 Görlitz, Germany}
\affiliation{Helmholtz-Zentrum Dresden-Rossendorf (HZDR), Bautzner Landstraße 400, 01328 Dresden, Germany}
\affiliation{Institut für Physik, Universität Rostock, D-18057 Rostock, Germany}
\author{Anton Reinhard}
\affiliation{Center for Advanced Systems Understanding (CASUS), Untermarkt 20, 02826 Görlitz, Germany}
\affiliation{Helmholtz-Zentrum Dresden-Rossendorf (HZDR), Bautzner Landstraße 400, 01328 Dresden, Germany}
\author{Simeon Ehrig}
\affiliation{Center for Advanced Systems Understanding (CASUS), Untermarkt 20, 02826 Görlitz, Germany}
\affiliation{Helmholtz-Zentrum Dresden-Rossendorf (HZDR), Bautzner Landstraße 400, 01328 Dresden, Germany}
\author{Klaus Steiniger}
\affiliation{Center for Advanced Systems Understanding (CASUS), Untermarkt 20, 02826 Görlitz, Germany}
\affiliation{Helmholtz-Zentrum Dresden-Rossendorf (HZDR), Bautzner Landstraße 400, 01328 Dresden, Germany}
\author{Michael Bussmann}
\affiliation{Center for Advanced Systems Understanding (CASUS), Untermarkt 20, 02826 Görlitz, Germany}
\affiliation{Helmholtz-Zentrum Dresden-Rossendorf (HZDR), Bautzner Landstraße 400, 01328 Dresden, Germany}
\author{Tobias Dornheim}
\affiliation{Helmholtz-Zentrum Dresden-Rossendorf (HZDR), Bautzner Landstraße 400, 01328 Dresden, Germany}
\affiliation{Center for Advanced Systems Understanding (CASUS), Untermarkt 20, 02826 Görlitz, Germany}


\date{\today}

\begin{abstract}
A key diagnostic in warm-dense matter (WDM) experiments is X-ray Thomson scattering (XRTS), but its interpretation is often limited by complex instrument effects and the high computationally expensive combinations of microscopic models with detector simulations. We present a proof-of-principle implementation of an event-driven approach to XRTS modelling, inspired by particle physics event-generators. Instead of computing the spectra via forward models, individual scattering events are sampled from the differential cross section and sent through a spectrometer simulation. This provides a statistically consistent representation that preserves full kinematic information and enables flexible and geometry-aware analysis. We demonstrate the feasibility and physical consistency of the method for non-resonant XRTS in a synthetic setup. By decoupling event generation from detector-level analysis, the framework allows efficient reuse of the sampled events and reduces computational overhead associated with repeated evaluations. The method is model-agnostic and establishes a new connection between particle-physics event generation techniques and WDM diagnostics, providing a scalable foundation for advanced XRTS analysis and inference.
\end{abstract}

\maketitle


\section{Introduction}
X-ray Thomson scattering (XRTS) \cite{gregori2003theoretical, glenzer2009x, sheffield2010plasma, falk2018experimental} is a key diagnostic tool for experiments on matter under extreme conditions, where temperatures of several tens of electronvolts coexist with near-solid densities. Such warm-dense matter (WDM \cite{graziani2014frontiers, vorberger2025roadmap}) states occur in technological applications, most notably inertial confinement fusion (ICF \cite{betti2016inertial, hurricane2023physics, hurricane2024present}) experiments and naturally in planetary interiors \cite{chabrier2006dense,militzer2016understanding,helled2020understanding,saumon2022current}. Due to all of this interest, WDM is nowadays routinely generated at modern high-power laser and X-ray facilities. The high-energy density experiments enabled by these platforms include XRTS measurements on spacially structured and dynamically evolving targets, which are relevant for ICF and related applications. These experiments, however, face challenges associated with low scattered-photon counts, complex source and instrument responses, evolving sample conditions, and strong geometric effects \cite{Chapman_POP_2014, gawne2025strong}, all of which complicate the interpretation of the measured spectra.\\
Traditionally, XRTS analyses rely on forward modelling of the energy and momentum transfer using analytical or semi-analytical representations of the electron dynamic structure factor (DSF), which encodes the microscopic density response of the probed WDM state to an external perturbation. High-fidelity descriptions of the underlying physics are available, ranging from quantum many-body models \cite{glenzer2009x,gregori2003theoretical,Zhang_CPL_2024} to ab initio simulations based on path-integral Monte Carlo (PIMC \cite{Dornheim_NatComm_2025,dornheim_dynamic}) and time-dependent density functional theory (TDDFT \cite{dynamic2,Schoerner_PRE_2023,Moldabekov_MRE_2025,White_ElectronicStructure_2025}). These approaches provide accurate, up to formally exact, solutions of the DSF, which can be used as input for detector-level simulations such as ray-tracing pipelines.\\
In practice, however, the computational cost and complexity of high-fidelity ab initio approaches limit their direct use. While ray-tracing itself is computationally not expensive, it remains significantly more costly than simple convolution based approaches. Therefore, a repeated evaluation of the full simulation pipeline, as required for parameter scans and Bayesian inference, become computationally prohibitive when combining microscopic models with full detector simulations. \\
In this work, we present a proof of principle demonstrating that an event-driven approach can restructure this workflow by decoupling the expensive ab initio computations from subsequent detector-level analysis. Instead of directly computing the scattered spectrum, we sample from the differential cross section to generate individual scattering events (photons or rays), which are subsequently propagated through the detector simulation. This strategy is conceptually similar to event generators widely used in high-energy physics \cite{buckley2011general}, while remaining complementary to established particle-transport codes, such as MCNP \cite{forster2006mcnp}, Geant4 \cite{agostinelli2003geant4}, or OpenMC \cite{romano2013openmc}. Whereas such codes provide detailed geometry and particle transport, they typically lack a microscopic description of the plasma-specific scattering processes. The present approach focuses on incorporating the latter at the event level, with the perspective of extending towards more realistic geometries in further developments. \\
In the present study, we focus on the non-resonant part of the scattering signal and neglect any coherence effects. Furthermore, we restrict ourselves to a synthetic diagnostic setup, using default simulation parameters. The goal is not to provide a full analysis pipeline for experimental data, but rather to establish that the event-driven approach is technically feasible, physically consistent, and compatible with existing detector-simulation tools.\\
This methodological foundation opens opportunities for more sophisticated modelling approaches. While not implemented in the present work, the event-driven concept can be extended to include spatially varying sample properties, such as temperature and density gradients, which are increasingly relevant for modern XRTS platforms \cite{Golovkin_HEDP_2013,Chapman_POP_2014,Poole_PPCF_2025}.
Overall, the contributions of this work are
\begin{enumerate}
    \item the demonstration of a direct, event-based interface between electronic structure models and detector simulations, 
    \item an outline of a pathway toward a fully integrated XRTS modelling platform with statistical consistency, instrument awareness and extensibility to realistic WDM geometries; and
    \item the validation and benchmarking of the new simulation components.
\end{enumerate}
We expect this approach to complement existing analytical and convolution-based methods and to support future XRTS diagnostics at facilities such as the National Ignition Facility \cite{Tilo_Nature_2023,macdonald2023colliding} and HED-HiBEF at the European XFEL \cite{bespalov2026experimentalevidencebreakdownuniformelectrongas,gawne2024ultrahigh}.

\section{Materials and Methods}

\subsection{The XRTS Analysis Scheme}
\begin{figure}[ht]
    \centering
    \includegraphics[width=\linewidth]{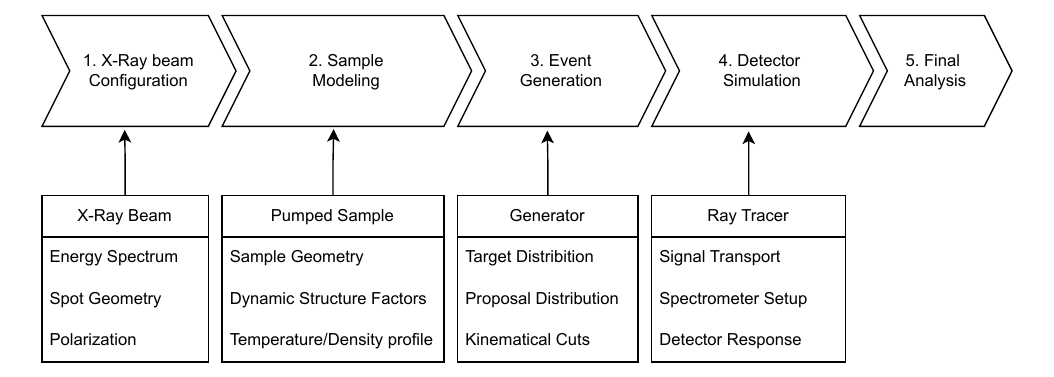}
    \caption{Schematic overview of the XRTS probing workflow, where the incident X-ray beam configuration and pumped-sample model are combined to generate scattering events imprinting the probing information, which are propagated through the detector simulation framework. Eventually, the synthetic detector image is compared to experimental data.}
    \label{fig:xrts_scheme}
\end{figure}
In a simplified picture, a typical pump-probe experiment for WDM consists of three main stages: 1) the pump stage, where a high-power optical laser deposits energy into a target creating a WDM state with temperature $T\gtrsim 10\,\mathrm{eV}$ and near-solid electron density. Such conditions are regularly  produced in major WDM campaigns, including the shock-compression XRTS measurements at LCLS \cite{nagler2015matter, glenzer2016matter}, the ICF-relevant WDM experiments at the National Ignition Facility \cite{kritcher2022design}, and, more recently, pump–probe studies at the HED-HiBEF endstation of the European XFEL \cite{bespalov2026experimentalevidencebreakdownuniformelectrongas}.
After a controlled delay, the pump stage is followed by 2) the probing stage, where x-rays are sent to be scattered/sent towards the pumped sample.\\
Finally, the probing is followed by 3) the detection stage, where the scattered photons are transported through x-ray capable optics and eventually counted in a dedicated x-ray spectrometer. For instance, at HED-HiBEF, a von-Hamos setup with a cylindrical mosaic crystal is commonly used to detect scattered photons with a high spectral resolution \cite{preston2020design,Gawne_JAP_2024}. \\
In Fig. \ref{fig:xrts_scheme}, an overview of the corresponding XRTS modelling workflow used in this work is depicted. The modelling pipeline begins with properties of the incident X-ray laser beam, such as spectral shape, focal-spot geometry and polarization, which are collected in a realistic representation. \\
In parallel, based on a temperature-density profile, a map of dynamic structure factors over the sample geometry is assembled to a realistic electronic structure model, where the profiles will be delivered by simulations performed for the pump stage, e.g., radiation-hydrodynamics (rad-hydro)\cite{zel2002physics,pomraning2005equations,mihalas2013foundations,Tilo_Nature_2023} or Particle-in-Cell simulations (PIC) \cite{Arber_2015,birdsall2018plasma}. The computation of the dynamic structure factor itself is obtained from (semi-)analytical models, ab initio calculations (such as TDDFT) or, potentially in the future, from analytic continuation procedures~\cite{Jarrell_PhysRep_1996,dornheim_dynamic,BENEDIXROBLES2026109904,chuna2025noiselesslimitimprovedpriorlimit}. These components together define the spatially resolved differential cross section of the scattering process.\\
With the beam and sample models established, the event-generator step samples from the differential cross section to produce individual scattering events, that are distributed in such a way, that the physical properties of the sample are encoded. Consequently, kinematic cuts given by the experimental configuration, such as angular constraints and spectral ranges of the detector setup, are applied during the generation. Instead of predicting the spectrum itself, a list of scattered photons with full momentum information is produced, which provides an event-level description of the probing stage. \\
The generated events are next propagated through a dedicated detector-simulation framework. Usually, a ray-tracer is used to simulate the spectrometer geometry, signal transport, crystal properties, and detector configuration, which converts the scattering events into a synthetic detector signal. This step includes instrument effects, such as angular acceptance, spectral dispersion and detector response, which are cumbersome to integrate into standard DSF-based forward models.\\
Finally, the detector-level output is processed to obtain observables that mimic the detector response of real-world experiments. The synthetic images and spectra generated by this workflow constitute an end-to-end simulation of the probing stage of a pump–probe experiment and enable both the benchmarking of the underlying physical models and the optimization of experimental design for specific aspects of the observed phenomena.

\subsection{Target Distribution: The Physical Model}
The central quantity in Monte-Carlo event generation for XRTS is the differential cross section, which encodes the local probability of a scattering event, where an X-ray photon with four-momentum $k_X$ scatters of an electron with four-momentum $p$ resulting in a scattered photon with four-momentum $k'$ and a recoil electron with four-momentum $p'$. In the formulation used here, the differential cross section serves as the target distribution to be sampled in the event generator and consists of three parts: the incident particle distribution $n(p_\mathrm{in})$, the in-medium modification $F(p_\mathrm{in},p_\mathrm{out})$ and the hard-scattering cross section $\mathrm{d}\sigma_{\mathrm{hard}}(p_{\mathrm{in}},p_{\mathrm{out}})$, which are combined into
\begin{align}\label{eq:general_diffCS}
    \mathrm{d}\sigma(p_{\mathrm{in}}, p_{\mathrm{out}}) = n(p_{\mathrm{in}})F(p_{\mathrm{in}},p_{\mathrm{out}})\mathrm{d}\sigma_{\mathrm{hard}}(p_{\mathrm{in}},p_{\mathrm{out}})\mathrm{d}p_{\mathrm{in}},
\end{align}
where we use the abbreviations $p_\mathrm{in} := (k_X, p)$ and $p_\mathrm{out} := (k', p')$.\\
The hard scattering cross section $\mathrm{d}\sigma_{\mathrm{hard}}(p_{\mathrm{in}},p_{\mathrm{out}})$ represents the raw scattering of a photon and an electron, without accounting for their embedding in surrounding matter. Since in typical pump-probe experiments, the photon energy of the X-ray probe is much smaller than the electron mass, $\omega_X\ll m_e$, the Thomson scattering cross section can be used:
\begin{align}\label{eq:hard_diffCS}
    \mathrm{d}\sigma_{\mathrm{hard}}(p_{\mathrm{in}},p_{\mathrm{out}}) = \mathrm{d}\sigma_{\mathrm{T}}(p_{\mathrm{in}},p_{\mathrm{out}}) = \frac{\alpha^2}{m^2}(\epsilon\cdot\epsilon')^2,
\end{align}
which encodes the angular and polarization dependence of the raw scattering event, where $\epsilon (\epsilon')$ is the polarization four-vector of the incoming (outgoing) photon.\\
The in-medium effects are introduced by the factor  
\begin{align}\label{eq:form_factor}
    F(p_{\mathrm{in}},p_{\mathrm{out}}) = \frac{\omega'}{\omega_X}S(Q),
\end{align}
where $\omega_X$ ($\omega'$) denotes the energy of the incoming (outgoing) photon, $Q:=k'-k_X=(\omega, \mathbf{q})$ the four-momentum transfer and $S(Q)\equiv S(\omega, \mathbf{q})$ the DSF, which describes density fluctuations of the plasma. Consequently, the DSF encodes how the surrounding matter modifies the energy and momentum transfer of the scattering process.\\
The final component of the differential cross section \eqref{eq:general_diffCS} is the four-momentum distribution of the incoming particles $n(p_\mathrm{in})\equiv n(k_X, p)$, which is given as
\begin{align}\label{eq:init_dist}
    n(k_X,p) = f_{X}(k_X),
\end{align}
where $f_X(k_X)$ denotes the X-ray photon four-momentum distribution and usually given by the X-ray beam model. For instance, assuming the photons of the X-ray beam all propagate along the $x$-axis and their energy is distributed according to the spectrum $f_X(\omega_X)$, the full photon distribution is given by\\
\begin{align}
    f_X(k_X) = f_X(\omega_X)\delta^{(3)}(\mathbf{\hat k}_X-\mathbf{e}_x),
\end{align}
where $\mathbf{\hat k}_X := \mathbf{k}_X/\abs{\mathbf{k}_X}$ denotes the normalized three-momentum of the X-ray photon, and $\mathbf{e}_x := (1,0,0)$.

\subsection{Sampling and Proposal}
To generate the events according to the differential cross section \eqref{eq:general_diffCS}, we use the standard acceptance-rejection algorithm \cite{VonNeumann1963, forsythe1972neumann,james1980monte,devroye2006nonuniform,lemieux2009monte,kroese2013handbook, rubinstein2016simulation}, which is schematically shown in Algorithm \ref{alg:rejection_sampler}. The generated events are samples drawn from a target distribution $F$, which represents the differential cross section, with the weight function $f(x = (p_\mathrm{in}, p_\mathrm{out})) = \mathrm{d}\sigma(p_\mathrm{in}, p_\mathrm{out})$. Since sampling from $F$ directly is usually not feasible, a proposal distribution $G$ with weight function $g(x)$ is introduced, that is assumed to be fast to sample. For each trial event $x_\mathrm{trial}\sim G$, a uniformly distributed probability $u\sim\mathcal{U}(0,1)$ is generated, and the event is accepted, if 
\begin{align}
    u < \frac{1}{w_\mathrm{max}} \frac{f(x_\mathrm{trial})}{g(x_\mathrm{trial})}, 
\end{align}
where $w_\mathrm{max}:=\sup_x w(x)$ is the maximum value of effective weight function $w(x)=\frac{f(x)}{g(x)}$. Loosely speaking, an event is accepted with a probability determined by the relative weight of the target distribution at the trial point, while the proposal distribution acts like a Jacobian of a coordinate transform that reshapes the sampling space to make the effective distribution flatter and easier to sample from. The accepted events are stored together with the residual weight $\max(1,w/w_\mathrm{max})$ to allow for slight over-weighting of events, which takes into account deviations between target and proposal distribution. This procedure is repeated until a desired number of events $N$ is accepted.\\
\begin{algorithm}
\caption{Acceptance-rejection algorithm to draw $N$ samples from a target distribution $F$ with weight function $f(x)$ using a proposal distribution $G$ with weight function $g(x)$.}\label{alg:rejection_sampler}
\DontPrintSemicolon
 \SetKwInOut{Input}{input}
\Input{ $N \geq 1$, target $F$, proposal $G$, maximum weight $w_\mathrm{max}$ }
\SetKwBlock{Beginn}{beginn}{ende}
\Begin{
    $n \gets 1$, $w \gets 0$

    $AccEvents \gets Vector(\mathrm{undef}, N)$

\While{$n \leq N$}{
    
    draw $x_\mathrm{trial}\sim G$
    
    $u\sim \mathcal{U}(0,1)$
    
    $w\gets f(x_\mathrm{trial})/g(x_\mathrm{trial})$
    
\uIf{$w > u\cdot w_{\mathrm{max}}$}{
    $AccEvents[i] \gets (x_{\mathrm{trial}},\max(1,w/w_\mathrm{max}))$\tcp*{accept the event}
    
    $n \gets n + 1$  
}
\Else{
    continue \tcp*{reject the event}
    }
}
}
\end{algorithm}
The mean number of trials needed to accept one event is called the efficiency of the acceptance-rejection algorithm. In more formal terms, this efficiency can be quantified by 
\begin{align}\label{eq:efficiency}
    \epsilon := \frac{\langle w_\mathrm{trial}\rangle}{w_\mathrm{max}}\quad \text{, where }\langle w_\mathrm{trial}\rangle = \left\langle \frac{f(x_\mathrm{trial})}{g(x_\mathrm{trial})}, \text{ with } x_\mathrm{trial}\sim G\right\rangle
\end{align}
is the expectation value of the weights at all trial points $x_\mathrm{trial}$ checked during the procedure. Consequently, the efficiency $\epsilon$ critically depends on the choice of the proposal distribution $G$. While a uniform proposal guarantees a complete coverage of the domain, it becomes very inefficient when the target distribution has prominent features like sharp peaks, long tails, or rapid variations. Therefore, $w(x)$ might become very large in localized regions, such that $w_\mathrm{max}$ strongly increases, while $ \langle w_\mathrm{trial}\rangle $ stays small, which leads to a low efficiency. \\
To address this issue, we optionally implemented a VEGAS-style proposal. The VEGAS algorithm \cite{Lepage:1977sw, Ohl:1998jn, Lepage:2020tgj} is well-known to increase the efficiency for sampling problems of not too large dimension. It adaptively constructs a step-function-like proposal using importance sampling by dividing the domain into bins, whose widths reflect the local magnitude of the effective weight function $w(x)$. In a training session, these bins are iteratively adjusted to adopt the strong features of $w(x)$. The resulting VEGAS-proposal can be sampled efficiently, and the samples are concentrated in regions of the domain, where the target distribution $f(x)$ is large. This reduces the variance of $w(x)$ and therefore significantly increases the efficiency \eqref{eq:efficiency}. \\
An important precomputed parameter of the acceptance-rejection algorithm and its efficiency is the maximum weight $w_\mathrm{max}:=\sup_x f(x)/g(x)$. If $w_\mathrm{max}$ is underestimated, the sampling process becomes biased, while if it is estimated too conservatively, the efficiency~\eqref{eq:efficiency} drops. Furthermore, finding the strict global maximum of $w(x)$ is often numerically unstable and prohibitively expensive. To balance these competing effects, we apply the quantile-reduction method (QR) \cite{Kish1995, danziger2022accelerating}, where one draws $N$ preliminary samples (usually with $N\sim 10^6$) together with their weights $\{w_i\}$ and sorts them, such that $w_i <= w_{i+1}$. Then the quantile-reduced maximum is given by 
\begin{align}
    w_\mathrm{max}^p := \min \left\{w_j \bigg\vert \sum_{i=j+1}^N w_i \leq p\sum_{i=1}^N w_i\right\},
\end{align}
where $p$ is the ratio of allowed over-weights, which are the largest weights which contribute only a fraction $p$ to the sum of all weights (usually $p\sim 10^{-3}$). This reduces the impact of rare outliers in $w(x)$ onto $w_\mathrm{max}$ and therefore is numerically more robust. Consequently, the quantile-reduced maximum $w_\mathrm{max}^p $ is per definition smaller than the actual global maximum and therefore increases the efficiency, while it is constructed in a way to minimize the bias introduced \cite{Kish1995}. 
\subsection{Signal Transport and Detector Simulation}\label{sec:ray_tracing}

For the fourth stage in the XRTS probing workflow shown in Fig.~\ref{fig:xrts_scheme}, the generated photon events are propagated through a realistic spectrometer setup using the High Energy Applications Ray Tracer (HEART; \cite{gawne2025heart}). Operating on an event-by-event basis, HEART takes the four-momenta of the outgoing photons and tracks their propagation through optical components and onto a pixelated detector. This allows for a direct interfacing between the event-driven XRTS modelling on the one hand and the simulation of the detector response of the other hand. Consequently, this approach does not rely on intermediate source-and-instrument functions. \\
In the present work we model a mosaic crystal spectrometer in von Hamos geometry, which is a common configuration in WDM experiments. In this geometry, a cylindrically bent crystal sits below the source-detector plane at its radius of curvature (ROC), and half way between the source and detector. This geometry focuses the x-rays into a line along the dispersive direction, making photon detections easier to observe above the noise level. We use a mosaic HAPG crystal as it is often used in experiments due to its balance/combination of high reflectivity and good resolution. Within HEART, photons are scattered according to dynamical diffraction theory, with reflectivity and absorption events calculated on each photons path.\\
HEART also accepts as input the four-momentum of all the sampled scattered photons. The ray tracing setup therefore provides a geometry-aware and event-based modelling of the spectrometer response, with a direct connection between the generated events and the synthetic detector image.

\subsection{Implementation}
The XRTS probing framework presented in this work is implemented using the Julia programming language \cite{Julia-2017} and wrapped in the package \texttt{XRTSprobing.jl} \cite{XRTSProbing}. This package provides the core functionality for the event generation, differential cross sections and includes the different variance-reduction and maximum-finding strategies discussed above. The used models to describe the in-medium modifications (in particular RPA) is provided by the package \texttt{ElectronicStructureModels.jl} \cite{ElectronicStructureModels}.  This modular separation between event generation and electronic structure modelling allows for straightforward substitution of alternative matter models, including first principles. \\
For fundamental data structures and kinematic operations, the implementation builds on the \texttt{QuantumElectrodynamics.jl} ecosystem \cite{QEDjl}. This includes standardized representations for four-momenta, particles and scattering processes, ensuring correctness and high performance.\\
The combination of these components result in a modular and extensible software stack, which allows for orthogonal development of the involved components such as microscopic physics models, event generation and detector-level simulations.  

\section{Results}
\subsection{Proof of Principle: Uniform Electron Gas}\label{sec:proof_of_principle}
To demonstrate the feasibility of the event-driven approach, we apply it to the uniform electron gas, which has been extensively studied in the literature \cite{review}.
The state of this system is characterized by the electron density $n_e$ and the temperature $T$, assuming a fully unpolarized system with $N_\uparrow = N_\downarrow = N/2$.
For a homogeneous electron gas, the characteristic momentum and energy scale\footnote{Unless stated otherwise, we employ natural units with $\hbar = c = k_B = 1$, such that the dimension of all quantities is expressed in terms of an energy scale, e.g., $[k_F] = [E_F] = [T] = [S(\omega, q)] = \mathrm{eV}$, and $[n_e] = \mathrm{eV}^3$.} are given by the Fermi wave-vector $k_F =\sqrt[3]{3\pi^2 n_e}$ and the Fermi energy $E_F = \frac{k_F^2}{2m_e}$, where $m_e$ denotes the electron mass. 
In this setting, the dynamic structure factor $S$, cf.~\eqref{eq:form_factor}, is connected to the dynamic density-density response function $\chi(Q) \equiv \chi(\omega, \mathbf{q})$ by the fluctuation-dissipation theorem\cite{kubo1966fluctuation}:
\begin{align}
    S(Q) = S(\omega, q) = \frac{-\Im [\chi(\omega, q)]}{n_e\pi(1-\exp(-\beta\omega))},
\end{align}
where $\beta = T^{-1}$ and $q = \abs{\mathbf{q}}$. Applying the random phase approximation (RPA), the interacting response function reads
\begin{align}
    \chi(\omega, q) = \frac{\chi_0(\omega, q)}{1-v_C(q)\chi_0(\omega,q)}
\end{align}
where $v_C(q) = \frac{e^2}{q^2}$ is the Coulomb potential in momentum space with the elementary charge\footnote{In natural units, the elementary charge is given by $e = \sqrt{4\pi\alpha}$ with the fine structure constant $\alpha \approx (137.035999177)^{-1}$ \cite{mohr2025codata}, while the vacuum permittivity reads $\varepsilon_0 = 1$.} $e$ and $\chi_0(\omega, q)$ is the finite-temperature Lindhard function \cite{giuliani2008quantum}, which describes the non-interacting electron gas. An explicit expression for $\chi_0(\omega, q)$ is given in Appendix \ref{appendix:Lindhard}.
For the probing configuration, we assume that all incident X-ray photons propagate along a common axis, say $\mathbf{e}_z = (0,0,1)$ and their energies follow a truncated Gaussian distribution:
\begin{align}
    f_X(\omega_X) \sim \exp\left(-\frac{(\omega_X-\omega_X^\mathrm{ref})^2}{\Delta\omega_X^2}\right)\Theta(\omega_X),
\end{align}
where $\omega_X^\mathrm{ref}$ denotes the central reference energy, $\Delta\omega_X$ the spectral width, and $\Theta(x)$ is the Heaviside step-function, which enforces the physical constraint $\omega_X > 0$. The incident electron energy is sampled uniformly, while the corresponding statistical weight is effectively absorbed in the dynamic structure factor. Together, these assumptions complete the setting for the differential XRTS cross section.
\begin{figure}[b]
    \centering
    \includegraphics[width=0.99\linewidth]{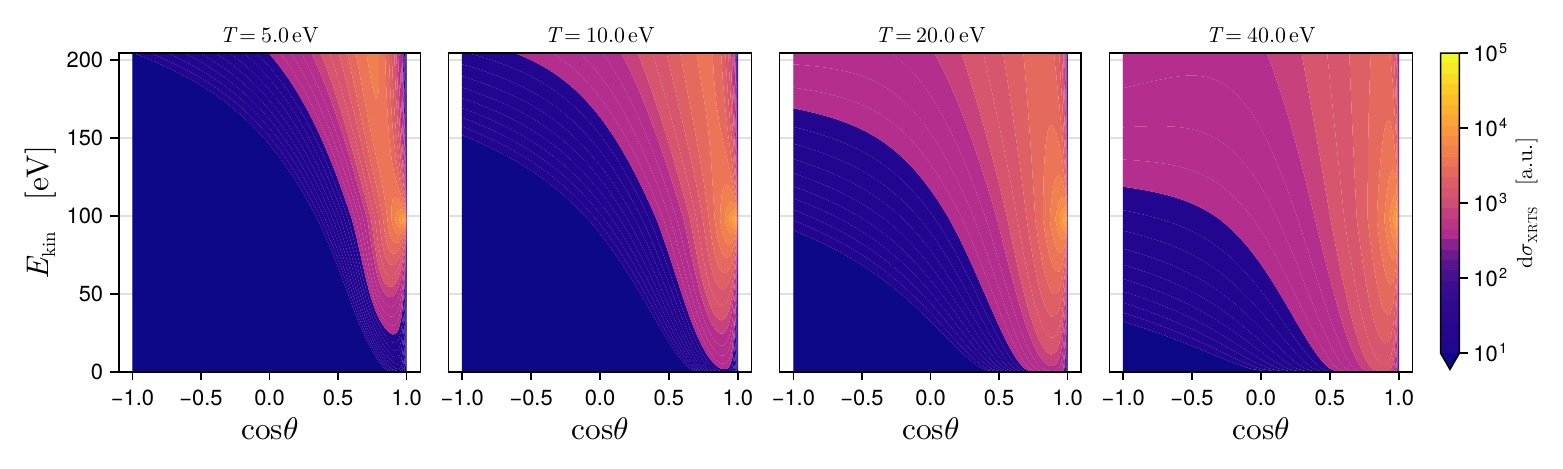}
    \caption{Differential cross section $\mathrm{d}\sigma(p_{\mathrm{in}}, p_{\mathrm{out}})$ for an interacting electron gas as a function of the kinetic energy $E_{\mathrm{kin}}$ of the incident electron and the cosine of the scattering angle $\cos\theta$, for the electron density $n_e = 10^{23}\mathrm{cm}^{-3}$ and electron temperatures of $T = 5, 10, 20,$ and $40\mathrm{eV}$. The energy of the incident photon is $\omega_X = 10\mathrm{keV}.$ }
    \label{fig:diffCS}
\end{figure}
Figure \ref{fig:diffCS} exhibits the resulting differential cross section \eqref{eq:general_diffCS} for the interacting electron gas as a function of the incident electron energy $E_{kin}$ and the cosine of the scattering angle $\cos\theta$ for several temperatures. At low temperatures, the scattering signal is strongly pronounced at the forward direction, i.e., $\cos\theta \to 1$. This reflects, as expected, the dominance of the collective and near-elastic scattering in degenerate electron gases. As the temperature increases, a redistribution of the spectrum towards larger angles and the contribution of higher electron energies can be observed, which is also expected due to thermal broadening. For the highest temperature $T=40\mathrm{eV}$, this results in a widely extended angular distribution, which indicates the reduced role of degeneracy effects and the increased importance of the single-particle scattering. This temperature dependence highlights the sensitivity of the differential cross section to the underlying electronic structure and provides therefore a suitable test case for the event-driven modelling approach. 

\subsection{Distribution of Scattered Events}
Building on the forward-model description introduced in Sec.~\ref{sec:proof_of_principle}, we generate $N = 10^6$ individual scattering events that carry the complete energy and momentum information of all incident and scattered particles. Subsequently, this event list is analysed through projections onto relevant kinematic components. 
\begin{figure}[t]
    \centering
    \includegraphics[width=0.99\linewidth]{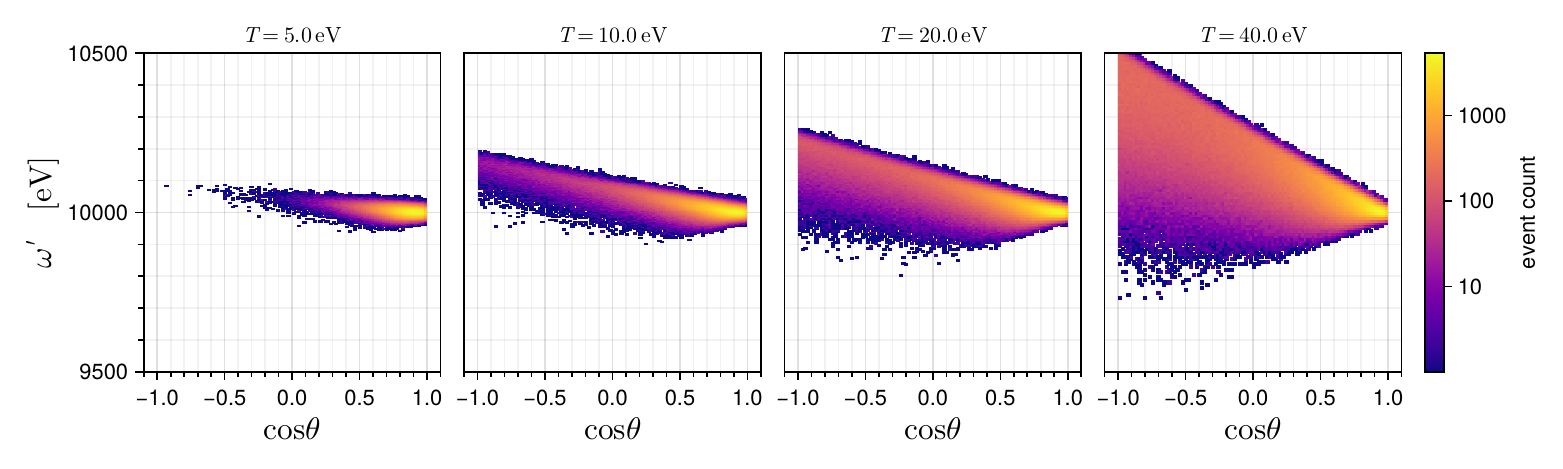}
    \caption{Generated events for XRTS off an interacting electron gas projected onto the plane span by the the cosine of the scattering angle $\cos\theta$ and the energy $\omega'$ of the scattered photon, for the electron density $n_e = 10^{23}\mathrm{cm}^{-3}$ and electron temperatures of $T = 5, 10, 20,$ and $40\mathrm{eV}$. The central energy of the incident photon is $\omega_X^\mathrm{ref} = 10\mathrm{keV}$ with a spectral width of $\Delta\omega_X = 0.1\, \mathrm{keV}$.}
    \label{fig:cth_omp}
\end{figure}
\begin{figure}[b]
    \centering
    \includegraphics[width=0.99\linewidth]{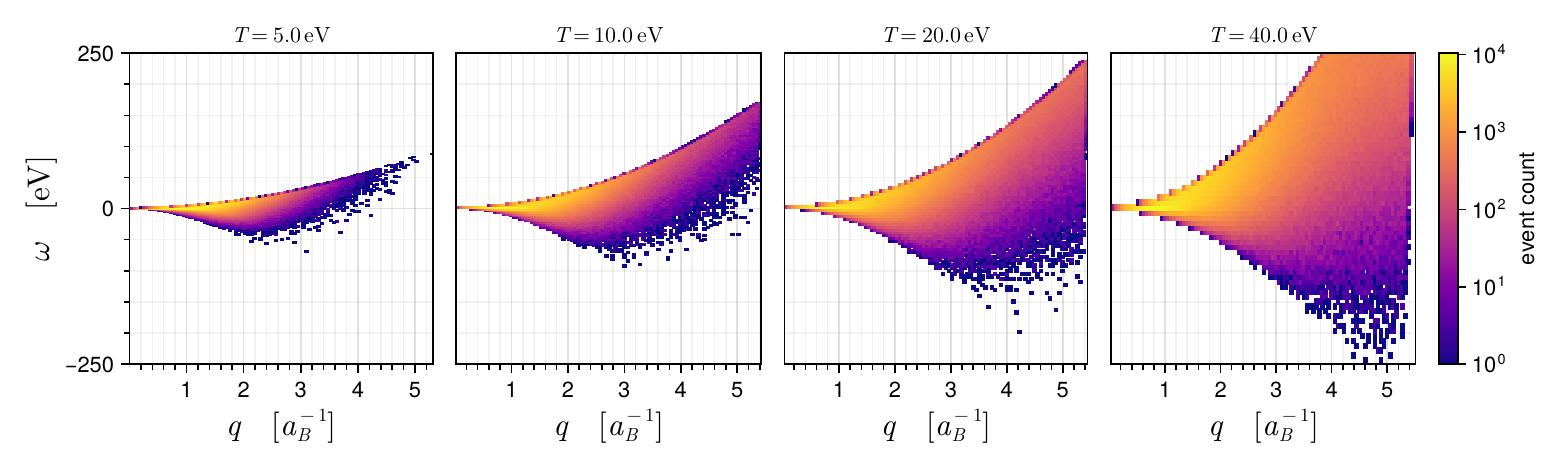}
     \caption{Same as in Fig.~\ref{fig:cth_omp}, but projected on the plane spanned by the energy transfer $\omega$ and momentum transfer $q$.}
    \label{fig:om_q}
\end{figure}
First, in Fig.~\ref{fig:cth_omp} we show the resulting 2D-histogram obtained from the generated event list, projected onto the plane spanned by the scattered photon energy $\omega'$ and the cosine of the scattering angle $\cos\theta$, which reflect the experimentally accessible observables and provide an intuitive view on the angular and energetic redistribution of the scattered photons. For low electron temperatures, the event distribution is centered around the forward direction, i.e., $\cos\theta \to 1$ and shows mostly an energy gain compared to the central X-ray energy of the initial photons of $\omega_X^\mathrm{ref} = 10\mathrm{keV}$. Rising temperatures result in significant broadening of the event distribution in both kinematic directions as well as a pronounced upswing towards energy-gain, which indicates thermal excitation and reduced electronic degeneracy. \\
Second, as displayed in Fig.~\ref{fig:om_q}, the same event list is projected onto the plane spanned by the energy transfer $\omega$ and momentum transfer $q$, where the sampled events directly trace the support and temperature dependence of the underlying electronic structure. As the temperature increases, the broadening from small energy transfer reflects the expanding phase space for single-particle contributions and the transition from a degenerate to a classical ideal gas. Similar to the observation made in Fig.~\ref{fig:cth_omp}, the energy-loss side (negative $\omega$) is exponentially suppressed relative to the energy-gain side (positive $\omega$), which is consistent with the detailed-balance property of the dynamic structure factor. Moreover, the distribution of events shows sharp kinematic boundaries characteristic of single-particle (on-shell) scattering. Off-shell contributions induced by in-medium effects are not resolved in this setup, and therefore collective phenomena, such as those associated with the plasma frequency, are not accessible. This simplified kinematic is chosen for clarity and serves as a minimal demonstration of the approach. \\
Together, these complementary projections demonstrate, that the generated events reproduce the structure of the conventional forward modelling, while providing a flexible discrete representation that can be interfaced directly with detector simulations and alternative analysis pipelines. 

\subsection{Validation}
\begin{figure}[ht]
    \centering
    \includegraphics[width=0.99\linewidth]{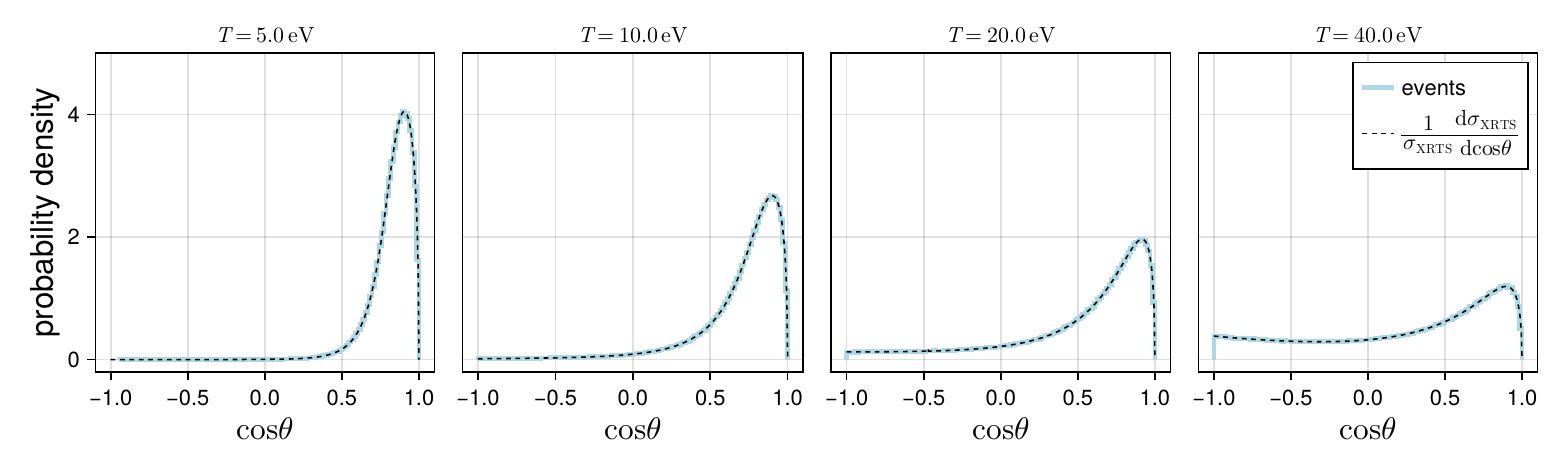}
    \caption{Same as in Fig.~\ref{fig:cth_omp}, but projected on the $\cos\theta$-axis. The black dashed line represents the normalized differential XRTS cross section \eqref{eq:general_diffCS} integrated over all kinematic degrees of freedom except $\cos\theta$.}
    \label{fig:compare}
\end{figure}
To check the consistency of the generated events with the forward model, we compare the angular distribution of the outgoing photon events with the differential cross section \eqref{eq:general_diffCS} integrated over all other kinematic degrees of freedom.\\
In Fig.~\ref{fig:compare} the probability density of the generated events is depicted along with the normalized XRTS cross section. For all temperatures shown, the generated events closely follow the computed ground-truth over the entire range of the scattering angle. In particular, the pronounced forward-scattering peak for low temperatures and the thermal broadening for increasing temperatures is accurately reproduced. This confirms that the sampling captures both the position and the normalization of the angular spectrum. \\
This comparison provides a systematic validation of the event-driven approach and demonstrates that the discrete events reproduce the expected marginal distribution of the continuous forward model. There is no need for reweighting or dedicated post-processing for this stage of the pipeline. 

\subsection{Detector-Level Results}

\begin{figure}[t]
    \centering
    \includegraphics[width=0.8\linewidth]{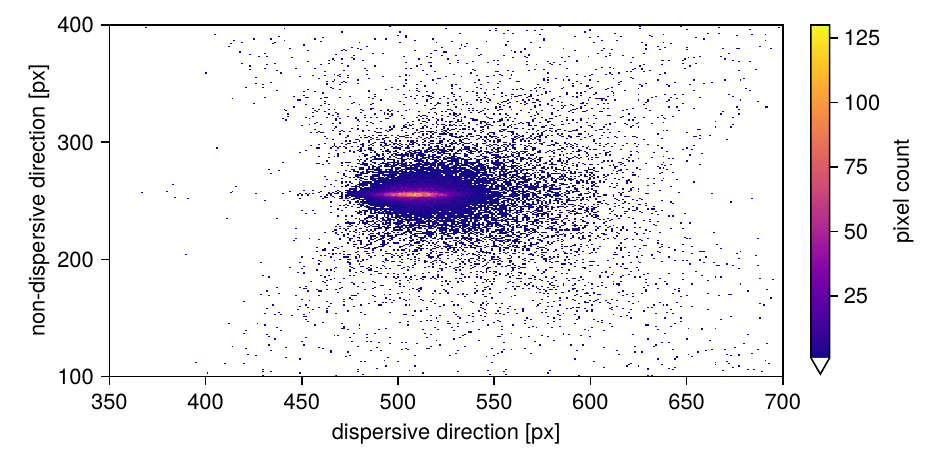}
    \caption{Detector image produced by HEART with a setup including a mosaic HAPG crystal with a radius of curvature $ROC = 80\,\mathrm{mm}$ and thickness $T_c = 0.04\,\mathrm{mm}$ combined with a Jungfrau detector. The covered window for the scattering angle is $\theta = 9.6^\circ - 11.75^\circ$. }
    \label{fig:detector}
\end{figure}
To demonstrate the full XRTS probing pipeline shown in Fig.~\ref{fig:xrts_scheme}, the generated events are passed over to the signal transport stage, where the outgoing photons are propagated from the scattering area to the detector using the ray-tracing code HEART (see Sec.~\ref{sec:ray_tracing}). At this point, the full four-momentum information is passed without intermediate binning or convolution. As the spectrometer setup, we use the von-Hamos geometry with a highly annealed pyrolytic graphite (HAPG) crystal \cite{preston2020design} and a Jungfrau detector \cite{mozzanica2016characterization}. 
Fig.~\ref{fig:detector} shows the resulting detector image for XRTS simulation of an electron gas with $n_e = 10^{23}\mathrm{cm}^{-3}$ and $T=5\mathrm{eV}$. The abscissa corresponds to the disperive direction of the setup, i.e. the axis along which the photon energy is mapped onto according to the crystal's dispersion relation. The detected signal is tightly concentrated around a central peak, which reflects the sharp spectral structure around the central frequency as already observed in Fig.~\ref{fig:cth_omp} (first panel). The ordinate represents the non-dispersive direction, where the detector exhibits a broad and defocused signal, which is typical for mosaic crystals\cite{sanchez1992conceptual, zastrau2012focal}. 
This demonstrates the consistency of the full event-driven XRTS probing workflow shown in Fig.~\ref{fig:xrts_scheme}: starting with the microscopic description of the electronic response encoded into the differential cross section, translated into scattering events via Monte-Carlo event generation and culminated into a realistic detector image by a dedicated ray-tracing code. This detector image represents a direct and geometry-aware realization of the XRTS signal without intermediate simplifications.

\subsection{Performance Analysis and Break-Even Considerations}
\begin{figure}[b]
    \centering
    \includegraphics[width=0.8\linewidth]{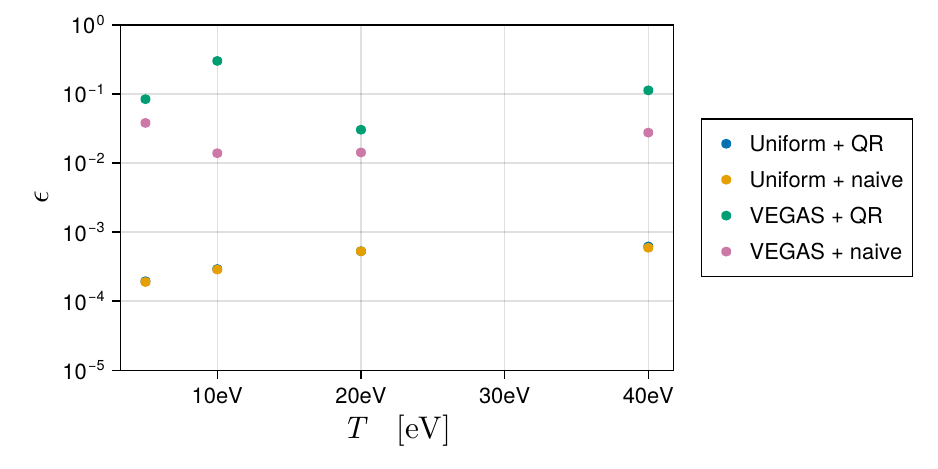}
    \caption{Efficiency of the acceptence-rejection algorithm \eqref{eq:efficiency} computed based on $10^7$ events generated with the same setup as used in Fig.~\ref{fig:cth_omp} and Fig.~\ref{fig:om_q} for different combinations of proposal distributions (uniform, VEGAS) and maximum finding strategies (naive, QR). }
    \label{fig:efficiency}
\end{figure}
Performance analysis of the event-generation stage is critical for its applicability, and therefore for the event-driven approach, especially when applied to produce large statistics mandatory for realistic experiment simulations.
In Fig.~\ref{fig:efficiency}, the efficiency $\epsilon$, for the sampling process of the XRTS cross section is depicted for different combinations of proposal distributions (Uniform, VEGAS) and maximum-finding strategies (naive, quantile reduction) at different electron temperatures. Each data point is acquired by sampling $10^7$ trial events and the efficiency is computed using Eq.~\eqref{eq:efficiency}. 
For uniform sampling, the efficiency lies in the range of $10^{-4} - 10^{-3}$ with a slight increase for higher temperatures, which can be traced back to the progressive smoothing of the distribution as shown in Figs.~\ref{fig:diffCS} and \ref{fig:cth_omp}. The application of QR to the uniform proposal shows no significant change of the efficiency, which indicates that the main limitation comes from the mismatch between proposal and target distribution. \\
In contrast, the application of the VEGAS proposal significantly improves the efficiency by orders of magnitude to the range $10^{-2}$. Combining VEGAS with QR even further improves the efficiency to $ \gtrsim 10^{-1}$. This shows, that the adaptive training of VEGAS correctly captures the peak structure of the target distribution, while QR further improves the rejection bound. Consequently, the combination of VEGAS and QR reaches the best performance over all considered temperatures. \\
In practice, the actual performance metric is the number of trial events $N_{\mathrm{trial}}$ which need to be generated in order to accept a desired number of events $N$. As a rule of thumb, for rejection sampling, the number of trials scale as $N_{\mathrm{trial}} = \frac{N}{\epsilon}$. For instance, for $N = 10^8$ and efficiencies $\epsilon_U = 10^{-3}$ (uniform) and $\epsilon_{VQR} = 10^{-1}$ (VEGAS + QR), the number of trial events yield $N_{\mathrm{trial}}^U = 10^{11}$ and $N_{\mathrm{trial}}^{VQR} = 10^{9}$, which corresponds to a reduction of two orders of magnitude for the number of generated events. \\
However, both, the VEGAS training and the QR estimate, need to generate some events upfront. Denoting the total number of overhead trial events $C_{VQR}$, the total number of trial events for VEGAS+QR is $T_{VQR} = C_{VQR} + \frac{N}{\epsilon_{VQR}}$.
The uniform sampling needs no training step, but the naive max-finding also comes with a number of overhead trail events $C_{\mathrm{naive}}$. Therefore, the total number of trail events for the combination of uniform sampling and naive max-finding yields $T_U = C_{\mathrm{naive}} + \frac{N}{\epsilon_U}$.
The break-even point of the condition $T_{VQR} \leq T_U$ is then given by 
\begin{align}
    N_{\mathrm{break}} = \frac{C_{VQR} - C_{\mathrm{naive}}}{\epsilon_{U}^{-1} - \epsilon_{VQR}^{-1}}.
\end{align}
For the realistic values $\epsilon_U = 10^{-3}$, $\epsilon_{VQR} = 10^{-1}$, $C_{VQR} = 2\cdot 10^{6}$ and $C_{\mathrm{naive}} = 10^6$, one obtains $N_{\mathrm{break}} \approx 10^3$. Consequently, even for moderately large sample sizes, the application of VEGAS and QR is beneficial for an effective sampling process.\\
Finally, considering that the generation of trial events using VEGAS is more expensive than uniform sampling, say by a factor for $r$, the break-even condition reads $r\cdot C_{VQR} + \frac{rN}{\epsilon_{VQR}} \leq C_{\mathrm{naive}} + \frac{N}{\epsilon_U}$. Thus, the number of events for a break-even scales linearly with $r$. Realistically, $r$ is on the order of 10, which increases the number of desired accepted events for break-even to $N_{\mathrm{break}} \approx 10^4$. Therefore, even under conservative assumptions, the variance-reduction techniques and maximum reduction deployed here remain beneficial for the number of events relevant for realistic XRTS probing simulations. 

\section{Discussion and Conclusion}

In the present work, we examine an event-driven approach for the simulation pipeline of XRTS probing experiments. We presented a proof-of-principle implementation, bridging between the microscopic electronic response models and the simulation of the spectrometer. This approach is complementary to the traditional forward modelling by generating statistically consistent ensemble of scattering events that carry all kinematic information imprinted by the microscopic description. This ensemble can then be used in several different downstream analysis pipelines, including kinematic projections and full-fledged detector simulations. In practice, the computationally expensive step of sampling the differential cross section only has to be performed once.\\
This makes event-level modelling of the probing stage more flexible than conventional forward-modelling and convolution-based approaches. The latter usually uses marginalized quantities such as energy-integrated cross sections, and typically incorporates detector effects via convolution with simple source-and-instrument functions. In contrast, the event-driven approach preserves the full phase-space information, which enables geometry-aware detector simulations and consistent modelling of counting statistics within a unified framework. \\
Moreover, the event-driven approach has important implications for uncertainty quantification and Bayesian inference of experimental parameters~\cite{Kasim_POP_2019}. Since the method directly generates distributed events, likelihood functions can be directly constructed on event or detector pixel counts, which opens a direct pathway toward an inference framework that consistently includes microscopic models, uncertainties, and instrument effects. \\
In the current form, the proof-of-principle implementation has potential for improvements. Multi-photon effects, coherent scattering contributions and collective strong-field effects are not included yet, but are conceivable extensions to the existing code. Furthermore, the addition of off-shell contributions to capture collective effects in phase-space is planned for future extensions. Moreover, the assumption of statistical independence of the generated events limits the approach to the weak-scattering regime, which is similar to the assumptions made for linear response theory. Mitigating these assumptions is more challenging, and includes the modelling of the modification of the probed matter during the X-ray scattering process. In this context, for instance, particle-in-cell (PIC) methods are well suited for describing these strongly driven plasma dynamics. However, combining the strength of both, PIC-based methods and the presented event-driven scattering approach requires more methodological development and is therefore dedicated to future work.\\
The key strength of the framework presented here is the direct interfacing to first-principle descriptions in form of the dynamic structure factor. Therefore, the event generation is agnostic to the underlying microscopic model and demands only a commutable expression of the differential cross section. This enables the straightforward  integration of different models for response functions, for instance from TDDFT, PIMC or other many-body approaches. In this sense, the framework provides a modular bridge between the ab initio electronic structure models and the realistic experiment simulation. \\
In summary, we demonstrated the feasibility, physical consistency and computational efficiency of an event-driven approach to simulate the XRTS probing pipeline. Here, variance and maximum reduction methods enable high performance for the central sampling process, while the modular structure allows for straightforward inclusion of more advanced electronic structure models. The presented concept complements the well-established convolution-based approaches while offering more flexibility and natural inclusion of realistic detector simulations. Therefore, it provides a scalable foundation for the next-generation XRTS diagnostics simulation in HED science.

\begin{acknowledgments}
UHA sincerely thanks Thomas Chuna for his introduction to dynamic structure factors and very fruitful discussions.
This work was partially supported by the Center for Advanced Systems Understanding (CASUS) which is financed by Germany’s Federal Ministry of Education and Research (BMBF) and by the Saxon state government out of the State budget approved by the Saxon State Parliament.
Tobias Dornheim gratefully acknowledges funding from the Deutsche Forschungsgemeinschaft (DFG) via project DO 2670/1-1.
This work has received funding from the European Union's Just Transition Fund (JTF) within the project \emph{R\"ontgenlaser-Optimierung der Laserfusion} (ROLF), contract number 5086999001, co-financed by the Saxon state government out of the State budget approved by the Saxon State Parliament. 
This work has received funding from the European Research Council (ERC) under the European Union’s Horizon 2022 research and innovation programme (Grant agreement No. 101076233,“PREXTREME"). 
Views and opinions expressed are however those of the authors only and do not necessarily reflect those of the European Union or the European Research Council Executive Agency. Neither the European Union nor the granting authority can be held responsible for them.
\end{acknowledgments}

\appendix

\section[\appendixname~\thesection]{Non-Interacting Electron Gas}\label{appendix:Lindhard}
For completeness, we briefly summaries the mathematical expressions employed within the present work. An ideal electron gas is fully described by its electron number density $n_e$ and electron temperature $T$. The corresponding dynamic density-density response function $\chi_0(\omega, q)$ is given by the so-called Lindhard function \cite{lindhard1953properties}, where $\omega$ and $q$ denote the energy and three-momentum transfer, respectively. For convenient notation, we introduce the following dimensionless quantities: 
\begin{align}
    \ombar = \frac{\omega}{E_F},\qquad \qbar = \frac{q}{k_F},\qquad  \betabar = \frac{E_F}{T}, \qquad \mubar = \frac{\mu}{E_F}, 
\end{align}
where $k_F$ denotes the Fermi wave vector, $E_F$ the Fermi energy and $\mu$ the chemical potential. The latter is determined by the condition
\begin{align}
    \int\mathrm{d}x\,x^2 f(x,\betabar,\mubar) = \frac{1}{3},
\end{align}
where $f$ is the Fermi distribution given by
\begin{align}
    f(x,\betabar, \mubar) = \frac{1}{1+\exp[\betabar(x^2 - \mubar)]}.
\end{align}
At finite temperatures, the imaginary part of the Lindhard function reads \cite{giuliani2008quantum}
\begin{align}
    \Im[\chi_0(\omega, q)] &= - \frac{k_F}{2 \pi \qbar} \frac{1}{2\betabar \qbar} \ln\left[\frac{1+\exp(\betabar(\mubar - \nu_-^2))}{1+\exp(\betabar(\mubar - \nu_+^2))}\right], 
\end{align}
where $\nu_\pm = \frac{\ombar\pm \qbar^2}{2\qbar}$.
The corresponding real part of the Lindhard function can be written numerically stable as \cite{ancarani2016efficient}
\begin{align}
    \Re[\chi_0(\omega, q)] &= - \frac{k_F}{2 \pi \qbar} \left[g(\nu_-, \betabar) - g(\nu_+, \betabar)\right],
\end{align}
with the auxiliary function
\begin{align}
    g(\nu, \betabar) := \int_0^\infty\mathrm{d}y\, \frac{f'(\nu y, \betabar)}{2}\left(-y + \frac{1}{1-y^2}\ln\abs{\frac{1+y}{1-y}}\right).
\end{align}


\bibliographystyle{vancouver}
\bibliography{bibliography} 

@article{gregori2003theoretical,
  title={Theoretical model of x-ray scattering as a dense matter probe},
  author={Gregori, G and Glenzer, Siegfried H and Rozmus, W and Lee, RW and Landen, OL},
  journal={Physical Review E},
  volume={67},
  number={2},
  pages={026412},
  year={2003},
  publisher={APS}
}

@Article{Dornheim_NatComm_2025,
author={Dornheim, Tobias
and D{\"o}ppner, Tilo
and Tolias, Panagiotis
and B{\"o}hme, Maximilian P.
and Fletcher, Luke B.
and Gawne, Thomas
and Graziani, Frank R.
and Kraus, Dominik
and MacDonald, Michael J.
and Moldabekov, Zhandos A.
and Schwalbe, Sebastian
and Gericke, Dirk O.
and Vorberger, Jan},
title={Unraveling electronic correlations in warm dense quantum plasmas},
journal={Nature Communications},
year={2025},
month={Jun},
day={02},
volume={16},
number={1},
pages={5103},
issn={2041-1723},
doi={10.1038/s41467-025-60278-3},
}

@article{Zhang_CPL_2024,
doi = {10.1088/0256-307X/41/1/017801},
year = {2024},
month = {jan},
publisher = {Chinese Physical Society and IOP Publishing Ltd},
volume = {41},
number = {1},
pages = {017801},
author = {Zhang, Yupei and Mo, Chongjie and Zhang, Ping and Kang, Wei},
title = {A Composite Ansatz for Calculation of Dynamical Structure Factor},
journal = {Chinese Physics Letters}
}

@article{glenzer2009x,
  title={X-ray Thomson scattering in high energy density plasmas},
  author={Glenzer, Siegfried H and Redmer, Ronald},
  journal={Reviews of Modern Physics},
  volume={81},
  number={4},
  pages={1625--1663},
  year={2009},
  publisher={APS}
}

@book{sheffield2010plasma,
  title={Plasma scattering of electromagnetic radiation: theory and measurement techniques},
  author={Sheffield, John and Froula, Dustin and Glenzer, Siegfried H and Luhmann Jr, Neville C},
  year={2010},
  publisher={Academic press}
}

@article{falk2018experimental,
  title={Experimental methods for warm dense matter research},
  author={Falk, Katerina},
  journal={High Power Laser Science and Engineering},
  volume={6},
  pages={e59},
  year={2018},
  publisher={Cambridge University Press}
}

@book{graziani2014frontiers,
  title={Frontiers and challenges in warm dense matter},
  author={Graziani, Frank and Desjarlais, Michael P and Redmer, Ronald and Trickey, Samuel B},
  volume={96},
  year={2014},
  publisher={Springer Science \& Business}
}

@article{vorberger2025roadmap,
  title={Roadmap for warm dense matter physics},
  author={Vorberger, Jan and Graziani, Frank and Riley, David and Baczewski, Andrew D and Baraffe, Isabelle and Bethkenhagen, Mandy and Blouin, Simon and B{\"o}hme, Maximilian P and Bonitz, Michael and Bussmann, Michael and others},
  journal={arXiv preprint arXiv:2505.02494},
  year={2025}
}

@article{betti2016inertial,
  title={Inertial-confinement fusion with lasers},
  author={Betti, R and Hurricane, OA},
  journal={Nature Physics},
  volume={12},
  number={5},
  pages={435--448},
  year={2016},
  publisher={Nature Publishing Group UK London}
}

@article{hurricane2023physics,
  title={Physics principles of inertial confinement fusion and US program overview},
  author={Hurricane, Omar A and Patel, PK and Betti, R and Froula, Dustin H and Regan, SP and Slutz, Stephen A and Gomez, MR and Sweeney, Mary Ann},
  journal={Reviews of Modern Physics},
  volume={95},
  number={2},
  pages={025005},
  year={2023},
  publisher={APS}
}

@article{hurricane2024present,
  title={Present understanding of ignition and gain using indirect-drive inertial confinement fusion target designs on the US National Ignition Facility},
  author={Hurricane, OA and Allen, A and Bachmann, BL and Baker, KL and Baxamusa, S and Bhandarkar, SD and Biener, J and Bionta, SRM and Braun, T and Briggs, T and others},
  journal={Plasma Physics and Controlled Fusion},
  volume={67},
  number={1},
  pages={015019},
  year={2024},
  publisher={IOP Publishing}
}

@article{chabrier2006dense,
  title={Dense plasmas in astrophysics: from giant planets to neutron stars},
  author={Chabrier, Gilles and Saumon, D and Potekhin, AY},
  journal={Journal of Physics A: Mathematical and General},
  volume={39},
  number={17},
  pages={4411},
  year={2006},
  publisher={IOP Publishing}
}

@article{militzer2016understanding,
  title={Understanding Jupiter's interior},
  author={Militzer, Burkhard and Soubiran, Fran{\c{c}}ois and Wahl, Sean M and Hubbard, William},
  journal={Journal of Geophysical Research: Planets},
  volume={121},
  number={9},
  pages={1552--1572},
  year={2016},
  publisher={Wiley Online Library}
}

@article{helled2020understanding,
  title={Understanding dense hydrogen at planetary conditions},
  author={Helled, Ravit and Mazzola, Guglielmo and Redmer, Ronald},
  journal={Nature Reviews Physics},
  volume={2},
  number={10},
  pages={562--574},
  year={2020},
  publisher={Nature Publishing Group UK London}
}

@article{saumon2022current,
  title={Current challenges in the physics of white dwarf stars},
  author={Saumon, Didier and Blouin, Simon and Tremblay, Pier-Emmanuel},
  journal={Physics Reports},
  volume={988},
  pages={1--63},
  year={2022},
  publisher={Elsevier}
}

@article{gawne2025strong,
  title={Strong geometry dependence of the x-ray Thomson scattering spectrum in single crystal silicon},
  author={Gawne, Thomas and Moldabekov, Zhandos A and Humphries, Oliver S and Appel, Karen and Baehtz, Carsten and Bouffetier, Victorien and Brambrink, Erik and Cangi, Attila and Cr{\'e}pisson, Celine and G{\"o}de, Sebastian and others},
  journal={Electronic Structure},
  volume={7},
  number={2},
  pages={025002},
  year={2025},
  publisher={IOP Publishing}
}

@article{Chapman_POP_2014,
    author = {Chapman, D. A. and Kraus, D. and Kritcher, A. L. and Bachmann, B. and Collins, G. W. and Falcone, R. W. and Gaffney, J. A. and Gericke, D. O. and Glenzer, S. H. and Guymer, T. M. and Hawreliak, J. A. and Landen, O. L. and Le Pape, S. and Ma, T. and Neumayer, P. and Nilsen, J. and Pak, A. and Redmer, R. and Swift, D. C. and Vorberger, J. and Döppner, T.},
    title = {Simulating x-ray Thomson scattering signals from high-density, millimetre-scale plasmas at the National Ignition Facility},
    journal = {Physics of Plasmas},
    volume = {21},
    number = {8},
    pages = {082709},
    year = {2014},
    month = {08},
    issn = {1070-664X},
    doi = {10.1063/1.4893146},
    eprint = {https://pubs.aip.org/aip/pop/article-pdf/doi/10.1063/1.4893146/12685882/082709_1_online.pdf},
}

@article{Jarrell_PhysRep_1996,
title = {Bayesian inference and the analytic continuation of imaginary-time quantum Monte Carlo data},
journal = {Physics Reports},
volume = {269},
number = {3},
pages = {133-195},
year = {1996},
issn = {0370-1573},
doi = {https://doi.org/10.1016/0370-1573(95)00074-7},
author = {Mark Jarrell and J.E. Gubernatis}
}

@misc{chuna2025noiselesslimitimprovedpriorlimit,
      title={The noiseless limit and improved-prior limit of the maximum entropy method and their implications for the analytic continuation problem}, 
      author={Thomas Chuna and Nicholas Barnfield and Paul Hamann and Sebastian Schwalbe and Michael P. Friedlander and Tobias Dornheim},
      year={2025},
      eprint={2511.06915},
      archivePrefix={arXiv},
      primaryClass={physics.comp-ph},
      url={https://arxiv.org/abs/2511.06915}, 
}

@article{Kasim_POP_2019,
    author = {Kasim, M. F. and Galligan, T. P. and Topp-Mugglestone, J. and Gregori, G. and Vinko, S. M.},
    title = {Inverse problem instabilities in large-scale modeling of matter in extreme conditions},
    journal = {Physics of Plasmas},
    volume = {26},
    number = {11},
    pages = {112706},
    year = {2019},
    month = {11},
    issn = {1070-664X},
    doi = {10.1063/1.5125979},
    eprint = {https://pubs.aip.org/aip/pop/article-pdf/doi/10.1063/1.5125979/12607940/112706_1_online.pdf},
}

@article{review,
author = {T. Dornheim and S. Groth and M. Bonitz},
journal = {Phys. Reports},
pages = {1-86},
title = {The uniform electron gas at warm dense matter conditions},
volume = {744},
year = {2018},
doi = {10.1016/j.physrep.2018.04.001}
}

@article{BENEDIXROBLES2026109904,
title = {PyLIT: Reformulation and implementation of the analytic continuation problem using kernel representation methods},
journal = {Computer Physics Communications},
volume = {319},
pages = {109904},
year = {2026},
issn = {0010-4655},
doi = {https://doi.org/10.1016/j.cpc.2025.109904},
author = {Alexander {Benedix Robles} and Phil-Alexander Hofmann and Thomas Chuna and Tobias Dornheim and Michael Hecht}
}

@article{Tilo_Nature_2023,
author={D{\"o}ppner, T.
and Bethkenhagen, M.
and Kraus, D.
and Neumayer, P.
and Chapman, D. A.
and Bachmann, B.
and Baggott, R. A.
and B{\"o}hme, M. P.
and Divol, L.
and Falcone, R. W.
and Fletcher, L. B.
and Landen, O. L.
and MacDonald, M. J.
and Saunders, A. M.
and Sch{\"o}rner, M.
and Sterne, P. A.
and Vorberger, J.
and Witte, B. B. L.
and Yi, A.
and Redmer, R.
and Glenzer, S. H.
and Gericke, D. O.},
title={Observing the onset of pressure-driven K-shell delocalization},
journal={Nature},
year={2023},
month={May},
day={24},
issn={1476-4687},
doi={10.1038/s41586-023-05996-8},
}

@article{Gawne_JAP_2024,
    author = {Gawne, Thomas and Bellenbaum, Hannah and Fletcher, Luke B. and Appel, Karen and Baehtz, Carsten and Bouffetier, Victorien and Brambrink, Erik and Brown, Danielle and Cangi, Attila and Descamps, Adrien and Goede, Sebastian and Hartley, Nicholas J. and Herbert, Marie-Luise and Hesselbach, Philipp and Höppner, Hauke and Humphries, Oliver S. and Konôpková, Zuzana and Laso Garcia, Alejandro and Lindqvist, Björn and Lütgert, Julian and MacDonald, Michael J. and Makita, Mikako and Martin, Willow and Mishchenko, Mikhail and Moldabekov, Zhandos A. and Nakatsutsumi, Motoaki and Naedler, Jean-Paul and Neumayer, Paul and Pelka, Alexander and Qu, Chongbing and Randolph, Lisa and Rips, Johannes and Toncian, Toma and Vorberger, Jan and Wollenweber, Lennart and Zastrau, Ulf and Kraus, Dominik and Preston, Thomas R. and Dornheim, Tobias},
    title = {Effects of mosaic crystal instrument functions on x-ray Thomson scattering diagnostics},
    journal = {Journal of Applied Physics},
    volume = {136},
    number = {10},
    pages = {105902},
    year = {2024},
    month = {09},
    issn = {0021-8979},
    doi = {10.1063/5.0222072},
    eprint = {https://pubs.aip.org/aip/jap/article-pdf/doi/10.1063/5.0222072/20150309/105902_1_5.0222072.pdf},
}

@article{buckley2011general,
  title={General-purpose event generators for LHC physics},
  author={Buckley, Andy and Butterworth, Jonathan and Gieseke, Stefan and Grellscheid, David and H{\"o}che, Stefan and Hoeth, Hendrik and Krauss, Frank and L{\"o}nnblad, Leif and Nurse, Emily and Richardson, Peter and others},
  journal={Physics Reports},
  volume={504},
  number={5},
  pages={145--233},
  year={2011},
  publisher={Elsevier}
}

@inproceedings{forster2006mcnp,
  title={MCNP-a general Monte Carlo code for neutron and photon transport},
  author={Forster, RA and Godfrey, TNK},
  booktitle={Monte-Carlo Methods and Applications in Neutronics, Photonics and Statistical Physics: Proceedings of the Joint Los Alamos National Laboratory-Commissariat {\`a} l'Energie Atomique Meeting Held at Cadarache Castle, Provence, France April 22--26, 1985},
  pages={33--55},
  year={2006},
  organization={Springer}
}

@article{agostinelli2003geant4,
  title={Geant4—a simulation toolkit},
  author={Agostinelli, Sea and Allison, John and Amako, K al and Apostolakis, John and Araujo, Henrique and Arce, Pedro and Asai, Makoto and Axen, D and Banerjee, Swagato and Barrand, GJNI and others},
  journal={Nuclear instruments and methods in physics research section A: Accelerators, Spectrometers, Detectors and Associated Equipment},
  volume={506},
  number={3},
  pages={250--303},
  year={2003},
  publisher={Elsevier}
}

@article{romano2013openmc,
  title={The OpenMC monte carlo particle transport code},
  author={Romano, Paul K and Forget, Benoit},
  journal={Annals of Nuclear Energy},
  volume={51},
  pages={274--281},
  year={2013},
  publisher={Elsevier}
}

@article{nagler2015matter,
  title={The matter in extreme conditions instrument at the Linac coherent light source},
  author={Nagler, Bob and Arnold, Brice and Bouchard, Gary and Boyce, Richard F and Boyce, Richard M and Callen, Alice and Campell, Marc and Curiel, Ruben and Galtier, Eric and Garofoli, Justin and others},
  journal={Synchrotron Radiation},
  volume={22},
  number={3},
  pages={520--525},
  year={2015},
  publisher={International Union of Crystallography}
}

@article{glenzer2016matter,
  title={Matter under extreme conditions experiments at the Linac Coherent Light Source},
  author={Glenzer, SH and Fletcher, LB and Galtier, Eric and Nagler, Bob and Alonso-Mori, Roberto and Barbrel, B and Brown, SB and Chapman, DA and Chen, Zhijiang and Curry, CB and others},
  journal={Journal of Physics B: Atomic, Molecular and Optical Physics},
  volume={49},
  number={9},
  pages={092001},
  year={2016},
  publisher={IOP Publishing}
}

@article{kritcher2022design,
  title={Design of inertial fusion implosions reaching the burning plasma regime},
  author={Kritcher, AL and Young, CV and Robey, HF and Weber, CR and Zylstra, AB and Hurricane, OA and Callahan, DA and Ralph, JE and Ross, JS and Baker, KL and others},
  journal={Nature Physics},
  volume={18},
  number={3},
  pages={251--258},
  year={2022},
  publisher={Nature Publishing Group UK London}
}

@article{preston2020design,
  title={Design and performance characterisation of the HAPG von H{\'a}mos spectrometer at the high energy density instrument of the European XFEL},
  author={Preston, TR and G{\"o}de, S and Schwinkendorf, J-P and Appel, K and Brambrink, E and Cerantola, VALERIO and H{\"o}ppner, H and Makita, M and Pelka, Alexander and Prescher, C and others},
  journal={Journal of Instrumentation},
  volume={15},
  number={11},
  pages={P11033--P11033},
  year={2020}
}

@article{mozzanica2016characterization,
  title={Characterization results of the JUNGFRAU full scale readout ASIC},
  author={Mozzanica, A and Bergamaschi, A and Brueckner, M and Cartier, S and Dinapoli, R and Greiffenberg, D and Jungmann-Smith, J and Maliakal, D and Mezza, D and Ramilli, M and others},
  journal={Journal of Instrumentation},
  volume={11},
  number={02},
  pages={C02047--C02047},
  year={2016}
}

@book{pomraning2005equations,
  title={The equations of radiation hydrodynamics},
  author={Pomraning, Gerald C},
  year={2005},
  publisher={Courier Corporation}
}

@book{mihalas2013foundations,
  title={Foundations of radiation hydrodynamics},
  author={Mihalas, Dimitri and Mihalas, Barbara Weibel},
  year={2013},
  publisher={Courier Corporation}
}

@book{zel2002physics,
  title={Physics of shock waves and high-temperature hydrodynamic phenomena},
  author={Zel'Dovich, Ya B and Raizer, Yu P},
  year={2002},
  publisher={Courier Corporation}
}

@book{birdsall2018plasma,
  title={Plasma physics via computer simulation},
  author={Birdsall, Charles K and Langdon, A Bruce},
  year={2018},
  publisher={CRC press}
}

@article{Arber_2015,
year = {2015},
month = {sep},
publisher = {IOP Publishing},
volume = {57},
number = {11},
pages = {113001},
author = {Arber, T D and Bennett, K and Brady, C S and Lawrence-Douglas, A and Ramsay, M G and Sircombe, N J and Gillies, P and Evans, R G and Schmitz, H and Bell, A R and Ridgers, C P},
title = {Contemporary particle-in-cell approach to laser-plasma modelling},
journal = {Plasma Physics and Controlled Fusion},
}

@article{Lepage:2020tgj,
    author = "Lepage, G. Peter",
    title = "{Adaptive multidimensional integration: VEGAS enhanced}",
    eprint = "2009.05112",
    archivePrefix = "arXiv",
    primaryClass = "physics.comp-ph",
    journal = "J. Comput. Phys.",
    volume = "439",
    pages = "110386",
    year = "2021"
}

@article{Ohl:1998jn,
    author = "Ohl, Thorsten",
    title = "{Vegas revisited: Adaptive Monte Carlo integration beyond factorization}",
    eprint = "hep-ph/9806432",
    archivePrefix = "arXiv",
    reportNumber = "IKDA-98-15",
    journal = "Comput. Phys. Commun.",
    volume = "120",
    pages = "13--19",
    year = "1999"
}

@article{Lepage:1977sw,
    author = "Lepage, G. Peter",
    title = "{A New Algorithm for Adaptive Multidimensional Integration}",
    reportNumber = "SLAC-PUB-1839-REV, SLAC-PUB-1839",
    journal = "J. Comput. Phys.",
    volume = "27",
    pages = "192",
    year = "1978"
}

@article{VonNeumann1963,
  title={Various techniques used in connection with random digits},
  author={Von Neumann, John},
  journal={John von Neumann, Collected Works},
  volume={5},
  pages={768--770},
  year={1963},
  publisher={MacMillan New York, USA}
}

@article{forsythe1972neumann,
  title={Von Neumann’s comparison method for random sampling from the normal and other distributions},
  author={Forsythe, George E},
  journal={Mathematics of Computation},
  volume={26},
  number={120},
  pages={817--826},
  year={1972}
}

@article{james1980monte,
  title={Monte Carlo theory and practice},
  author={James, Frederick},
  journal={Reports on progress in Physics},
  volume={43},
  number={9},
  pages={1145},
  year={1980},
  publisher={IOP Publishing}
}

@article{devroye2006nonuniform,
  title={Nonuniform random variate generation},
  author={Devroye, Luc},
  journal={Handbooks in operations research and management science},
  volume={13},
  pages={83--121},
  year={2006},
  publisher={Elsevier}
}

@book{rubinstein2016simulation,
  title={Simulation and the Monte Carlo method},
  author={Rubinstein, Reuven Y and Kroese, Dirk P},
  year={2016},
  publisher={John Wiley \& Sons}
}

@book{kroese2013handbook,
  title={Handbook of monte carlo methods},
  author={Kroese, Dirk P and Taimre, Thomas and Botev, Zdravko I},
  year={2013},
  publisher={John Wiley \& Sons}
}

@book{lemieux2009monte,
  title={Monte carlo and quasi-monte carlo sampling},
  author={Lemieux, Christiane},
  year={2009},
  series={Springer Series in Statistics},
  publisher={Springer New York, NY}
}

@book{Kish1995,
  title={Survey Sampling},
  author={Kish, Leslie},
  year={1995},
  series={Wiley Classics Library},
  publisher={John Wiley \& Sons}
}

@article{gawne2025heart,
  title={HEART: A new X-ray tracing code for mosaic crystal spectrometers},
  author={Gawne, Thomas and Schwalbe, Sebastian and Chuna, Thomas and Acosta, Uwe Hernandez and Preston, Thomas R and Dornheim, Tobias},
  journal={Computer Physics Communications},
  pages={109878},
  year={2025},
  publisher={Elsevier}
}

@article{danziger2022accelerating,
  title={Accelerating Monte Carlo event generation--rejection sampling using neural network event-weight estimates},
  author={Danziger, Katharina and Jan{\ss}en, Timo and Schumann, Steffen and Siegert, Frank},
  journal={SciPost Physics},
  volume={12},
  number={5},
  pages={164},
  year={2022}
}

@article{mohr2025codata,
  title={CODATA recommended values of the fundamental physical constants: 2022},
  author={Mohr, Peter J and Newell, David B and Taylor, Barry N and Tiesinga, Eite},
  journal={Reviews of Modern Physics},
  volume={97},
  number={2},
  pages={025002},
  year={2025},
  publisher={APS}
}

@article{sanchez1992conceptual,
  title={A conceptual model for ray tracing calculations with mosaic crystals},
  author={Sanchez del Rio, M and Bernstorff, S and Savoia, A and Cerrina, F},
  journal={Review of scientific instruments},
  volume={63},
  number={1},
  pages={932--935},
  year={1992},
  publisher={American Institute of Physics}
}

@article{zastrau2012focal,
  title={Focal aberrations of large-aperture HOPG von-H{\`a}mos x-ray spectrometers},
  author={Zastrau, U and Brown, CRD and D{\"o}ppner, T and Glenzer, SH and Gregori, G and Lee, HJ and Marschner, H and Toleikis, S and Wehrhan, O and F{\"o}rster, E},
  journal={Journal of Instrumentation},
  volume={7},
  number={09},
  pages={P09015--P09015},
  year={2012}
}

@article{lindhard1953properties,
  title={On the properties of a gas of charged particles},
  author={Lindhard, Jens},
  journal={Kgl. Danske Videnskab. Selskab Mat.-Fys. Medd.},
  volume={28},
  year={1953},
  publisher={Univ. of Copenhagen, Denmark}
}

@article{dynamic2,
author = {A. D. Baczewski and L. Shulenburger and M. P. Desjarlais and S. B. Hansen and R. J. Magyar},
journal = {Phys. Rev. Lett},
pages = {115004},
title = {X-ray Thomson Scattering in Warm Dense Matter without the Chihara Decomposition},
volume = {116},
year = {2016},
}

@article{Poole_PPCF_2025,
doi = {10.1088/1361-6587/ad9e74},
year = {2024},
month = {dec},
publisher = {IOP Publishing},
volume = {67},
number = {1},
pages = {015034},
author = {Poole, H and Cao, D and Epstein, R and Golovkin, I and Goncharov, V N and Hu, S X and Kasim, M and Vinko, S M and Walton, T and Regan, S P and Gregori, G},
title = {Investigating the impact of intermediate-mode perturbations on diagnosing plasma conditions in DT cryogenic implosions via synthetic x-ray Thomson scattering},
journal = {Plasma Physics and Controlled Fusion}
}

@misc{bespalov2026experimentalevidencebreakdownuniformelectrongas,
      title={Experimental Evidence for the Breakdown of Uniform-Electron-Gas Models in Warm Dense Aluminium}, 
      author={Dmitrii S. Bespalov and Ulf Zastrau and Zhandos A. Moldabekov and Thomas Gawne and Tobias Dornheim and Moyassar Meshhal and Alexis Amouretti and Michal Andrzejewski and Karen Appel and Carsten Baehtz and Erik Brambrink and Khachiwan Buakor and Carolina Camarda and David Chin and Gilbert Collins and Céline Crépeisson and Adrien Descamps and Jon Eggert and Luke Fletcher and Alessandro Forte and Gianluca Gregori and Marion Harmand and Oliver S. Humphries and Hauke Höppner and Jonas Kuhlke and William Lynn and Julian Lütgert and Masruri Masruri and Emma M. McBride and Ryan Stewart McWilliams and Alan Augusto Sanjuan Mora and Jean-Paul Naedler and Paul Neumayer and Charlotte Palmer and Alexander Pelka and Lea Pennacchioni and Calum Prestwood and Natalia A. Pukhareva and Chongbing Qu and Divyanshu Ranjan and Ronald Redmer and Michael Roper and Christoph Sahle and Samuel Schumacher and Jan-Patrick Schwinkendorf and Melanie J. Sieber and Madison Singleton and Ethan Smith and Christian Sternemann and Thomas Stevens and Michael Stevenson and Cornelius Strohm and Minxue Tang and Monika Toncian and Toma Toncian and Thomas Tschentscher and Sam M. Vinko and Justin S. Wark and Max Wilke and Dominik Kraus and Thomas R. Preston},
      year={2026},
      eprint={2509.10107},
      archivePrefix={arXiv},
      primaryClass={physics.plasm-ph},
      url={https://arxiv.org/abs/2509.10107}
}

@article{Golovkin_HEDP_2013,
title = {Simulation of X-ray scattering diagnostics in multi-dimensional plasma},
journal = {High Energy Density Physics},
volume = {9},
number = {3},
pages = {510-515},
year = {2013},
issn = {1574-1818},
doi = {https://doi.org/10.1016/j.hedp.2013.05.001},
author = {Igor Golovkin and Joseph J. MacFarlane and Pamela Woodruff and Iain Hall and Gianluca Gregori and James Bailey and Eric Harding and Tom Ao and Siegfried Glenzer}
}

@article{White_ElectronicStructure_2025,
doi = {10.1088/2516-1075/adad24},
year = {2025},
month = {feb},
publisher = {IOP Publishing},
volume = {7},
number = {1},
pages = {014001},
author = {White, Alexander J},
title = {Dynamical structure factors of warm dense matter from time-dependent orbital-free and mixed-stochastic-deterministic density functional theory},
journal = {Electronic Structure}
}

@article{Moldabekov_MRE_2025,
    author = {Moldabekov, Zhandos A. and Schwalbe, Sebastian and Gawne, Thomas and Preston, Thomas R. and Vorberger, Jan and Dornheim, Tobias},
    title = {Applying the Liouville–Lanczos method of time-dependent density-functional theory to warm dense matter},
    journal = {Matter and Radiation at Extremes},
    volume = {10},
    number = {4},
    pages = {047601},
    year = {2025},
    month = {05},
    issn = {2468-2047},
    doi = {10.1063/5.0263947},
    eprint = {https://pubs.aip.org/aip/mre/article-pdf/doi/10.1063/5.0263947/20510493/047601_1_5.0263947.pdf},
}

@article{Schoerner_PRE_2023,
  title = {X-ray Thomson scattering spectra from density functional theory molecular dynamics simulations based on a modified Chihara formula},
  author = {Sch\"orner, Maximilian and Bethkenhagen, Mandy and D\"oppner, Tilo and Kraus, Dominik and Fletcher, Luke B. and Glenzer, Siegfried H. and Redmer, Ronald},
  journal = {Phys. Rev. E},
  volume = {107},
  issue = {6},
  pages = {065207},
  numpages = {14},
  year = {2023},
  month = {Jun},
  publisher = {American Physical Society},
  doi = {10.1103/PhysRevE.107.065207},
}

@article{dornheim_dynamic,
author = {T. Dornheim and S. Groth and J. Vorberger and M. Bonitz},
journal = {Phys. Rev. Lett.},
title = {Ab initio Path Integral {M}onte {C}arlo Results for the Dynamic Structure Factor of Correlated Electrons: From the Electron Liquid to Warm Dense Matter},
volume = {121},
year = {2018},
doi={10.1103/PhysRevLett.121.255001},
pages = {255001},
}

@book{giuliani2008quantum,
  title={Quantum theory of the electron liquid},
  author={Giuliani, Gabriele and Vignale, Giovanni},
  year={2008},
  publisher={Cambridge university press}
}

@article{ancarani2016efficient,
  title={Efficient technique to evaluate the Lindhard dielectric function},
  author={Ancarani, Lorenzo Ugo and Jouin, Herve},
  journal={The European Physical Journal Plus},
  volume={131},
  number={4},
  pages={114},
  year={2016},
  publisher={Springer}
}

@article{kubo1966fluctuation,
  title={The fluctuation-dissipation theorem},
  author={Kubo, Rep},
  journal={Reports on progress in physics},
  volume={29},
  number={1},
  pages={255--284},
  year={1966}
}

@article{Julia-2017,
    title={Julia: A fresh approach to numerical computing},
    author={Bezanson, Jeff and Edelman, Alan and Karpinski, Stefan and Shah, Viral B},
    journal={SIAM {R}eview},
    volume={59},
    number={1},
    pages={65--98},
    year={2017},
    publisher={SIAM},
    doi={10.1137/141000671},
}

@misc{QEDjl,
	author = {Hernandez Acosta, Uwe and Reinhard, Anton and Ehrig, Simeon and Steiniger, Klaus and Bussmann, Michael},
	year = {2026-03-30},
	title = {{QuantumElectrodynamics.jl (Version v0.4.0)}},
	url = {https://github.com/QEDjl-project/QuantumElectrodynamics.jl},
	howpublished = {http://doi.org/10.14278/rodare.4584},
}

@misc{ElectronicStructureModels,
	author = {Hernandez Acosta, Uwe},
	year = {2025-12-19},
	title = {{ElectronicStructureModels.jl~(Version~v0.2.1)}},
	url = {https://github.com/JuliaXRTS/ElectronicStructureModels.jl},
}

@misc{XRTSProbing,
	author = {{Uwe Hernandez Acosta}},
	title = {{XRTSProbing.jl (Version~v0.0.1)}},
	year = {2026},
	url = {https://github.com/JuliaXRTS/XRTSProbing.jl},
}

@article{macdonald2023colliding,
  title={The colliding planar shocks platform to study warm dense matter at the National Ignition Facility},
  author={MacDonald, Mike J and Di Stefano, Carlos A and Doeppner, Tilo and Fletcher, LB and Flippo, Kirk Adler and Kalantar, Daniel and Merritt, Elizabeth Catherine and Ali, Suzanne J and Celliers, Peter M and Heredia, Ricardo and others},
  journal={Physics of Plasmas},
  volume={30},
  number={6},
  year={2023},
  publisher={AIP Publishing}
}

@article{gawne2024ultrahigh,
  title={Ultrahigh resolution x-ray Thomson scattering measurements at the European X-ray Free Electron Laser},
  author={Gawne, Thomas and Moldabekov, Zhandos A and Humphries, Oliver S and Appel, Karen and Baehtz, Carsten and Bouffetier, Victorien and Brambrink, Erik and Cangi, Attila and G{\"o}de, Sebastian and Kon{\^o}pkov{\'a}, Zuzana and others},
  journal={Physical Review B},
  volume={109},
  number={24},
  pages={L241112},
  year={2024},
  publisher={APS}
}

\end{document}